\documentclass[12pt]{article}
\pdfoutput=1 

\usepackage{jheppub} 

\usepackage[T1]{fontenc} 
\usepackage{amsmath,physics}
\usepackage{comment}
\usepackage{bbm}
\usepackage[percent]{overpic}
\usepackage[toc,page]{appendix}
\usepackage{etoolbox}
\usepackage{tabularx}

\BeforeBeginEnvironment{appendices}{\clearpage}
\AfterEndEnvironment{appendices}{\clearpage}
\AfterEndEnvironment{appendices}{\pretocmd{\section}{\clearpage}{}{}}{}
\AfterEndEnvironment{appendices}{\pretocmd{\subsubsection}{\clearpage}{}{}}{}

\usepackage{graphicx,float,caption,subcaption,xcolor}
\usepackage{hyperref}
\hypersetup{
    colorlinks=true,
    linkcolor=blue,
    filecolor=magenta,      
    urlcolor=cyan,
    pdftitle={Aligned Fields Double Copy to Kerr-NUT-(A)dS},
    pdfpagemode=FullScreen,
    }

\graphicspath{{./img/}} 
\usepackage{multirow}

\usepackage{bbm}

\notoc

\title{Aligned Fields Double Copy to Kerr-NUT-(A)dS}
\author[a]{Samarth Chawla}
\author[a]{and Cynthia Keeler}

\affiliation[a]{Department of Physics, Arizona State University,
Tempe, AZ 85281, USA}

\emailAdd{samarthc@asu.edu}
\emailAdd{keelerc@asu.edu}

\abstract{We find Abelian gauge fields that double copy to a large class of black hole
    spacetimes with spherical horizon topology known as the Kerr-NUT-(A)dS family. Using a multi-Kerr-Schild prescription, we extend the previously-known double copy structure for arbitrarily rotating general dimension black holes, to include NUT charges
and an arbitrary cosmological constant. In all cases, these single copy gauge fields are `aligned fields', because their nonzero components align with the principal tensor which generates the Killing structure of the spacetime.
In five dimensions, we additionally derive the same single-copy field strengths via the Weyl double copy procedure.}

\begin{document} 
\maketitle
\flushbottom

\section{Introduction}

The classical double copy, first presented in \cite{Monteiro:2014cda}, maps classical Yang-Mills solutions to
classical gravity solutions via a `squaring' procedure.  This procedure is based on the color-kinematics duality
for amplitudes (see \cite{Travaglini:2022uwo,Bern:2019prr,Adamo:2022dcm}).  As in color-kinematics duality, the
classical gravitational solution is built from two copies of the classical Yang-Mills solution, so the Yang-Mills
solution is called the `single copy'.  Since the original formulation eight years ago, many other examples of the
classical double copy have been built
\cite{Luna:2015paa,Luna:2016hge,Monteiro:2018xev,Ridgway:2015fdl,Adamo:2017nia,Bahjat-Abbas:2017htu,Carrillo-Gonzalez:2017iyj,Ilderton:2018lsf,Gurses:2018ckx,CarrilloGonzalez:2019gof,Lee:2018gxc,
Lescano:2020nve,Lescano:2021ooe,Bah:2019sda,Goldberger:2019xef,Kim:2019jwm,Luna:2020adi,Keeler:2020rcv,Easson:2020esh,Berman:2018hwd,Alkac:2021seh,Mkrtchyan:2022ulc,Adamo:2022rob,Alfonsi:2020lub,Adamo:2020qru,Gonzo:2021drq}. Recent reviews include \cite{Kosower:2022yvp} and chapter 7 of \cite{Bern:2019prr}.

The simplest example of the classical double copy starts with the Yang-Mills field arrangement due to a single color charge at the origin.  When the dilaton and two-form field are tuned to zero \cite{Kim:2019jwm,Luna:2020adi}, this field double-copies to the Schwarzschild black hole metric.  Indeed, the first classical double copy paper \cite{Monteiro:2014cda} proposes a single copy field strength which double copies to the broader class of Kerr black holes in higher dimensions: they include general-dimension black holes with arbitrary spins but flat asymptotics.

In this paper, we extend this result to include the full Kerr-NUT-(A)dS class of spacetimes \cite{Chen:2006xh}.  Our extension is possible because these spacetimes all possess a series of Killing objects (vectors and higher tensors), all related to the `principal tensor' of the spacetime (\cite{Kubiznak:2006kt}; for a recent review, see \cite{Frolov:2017kze}).  The principal tensor, a non-degenerate closed conformal Killing-Yano 2-form,  not only provides for separability due to the conserved quantities from the Killing objects (see  \cite{Frolov:2017kze} and citations therein, as well as \cite{Krtous:2018bvk}),
it also ensures the spacetime is algebraically special \cite{Mason:2010zzc}.  Indeed, the existence of the principal tensor, together with the assumption that the metric is a vacuum solution to Einstein's equation, uniquely identifies the Kerr-NUT-(A)dS class \cite{Krtous:2008tb}.

As we show, the single copy for the full class of these spacetimes is an `aligned field', so called because its components align with those of the principal tensor.  These aligned fields, found fully in \cite{Krtous:2007xg}, have charges which we set proportional to the mass and NUT charges of the spacetime.  We argue for our single copy proposal using a multi-Kerr-Schild ansatz.

Specifically in five dimensions, we additionally show that our proposed single-copy field matches the Weyl double copy.  In the Weyl double copy, a single copy field strength and zeroth copy scalar are proposed which combine to form the Weyl tensor directly.  Following the first proposal for Petrov type D spacetimes in \cite{Luna:2018dpt}, the Weyl double copy was extended to other Petrov types \cite{Godazgar:2020zbv} and has found justification from amplitudes directly \cite{Elor:2020nqe,Farnsworth:2021wvs,White:2020sfn,Chacon:2021wbr,Chacon:2021lox,Han:2022ubu,Easson:2021asd,Luna:2022dxo,Monteiro:2020plf}.  Aside from the spinor formalism (extended to five dimensions in \cite{Monteiro:2018xev}), we additionally verify our five-dimensional copy via the Weyl doubling formula of \cite{Alawadhi:2020jrv}.  For $D\geq 6$, we leave a thorough exploration of a Weyl double copy to future work.

In the next section, we review the classical double copy and the spinor formalism needed for its Weyl formulation, as well as the  Kerr-NUT-(A)dS spacetimes and their properties. In Section \ref{sec:FiveDDoubleCopy}, we construct the five-dimensional single copy using the Weyl spinor, and show its equivalence to the multi-Kerr-Schild ansatz.  We additionally note that this field is aligned to the principal tensor of the spacetime.  In Section \ref{sec:GeneralDSingleCopy}, we use a multi-Kerr-Schild approach, showing that the single copy of any Kerr-NUT-(A)dS spacetime is an aligned field.  Finally, in Section \ref{sec:discussion}, we discuss our results and directions for future work.  

\section{Review}\label{sec:review}

\subsection{Classical Double Copy}\label{sec:ReviewDC}
As we now review, the double copy is a map between the dynamics of gauge theory and gravity, where gravity is realized in terms of two copies of a gauge field. 
The earliest formulation \cite{Monteiro:2014cda} of this map takes as its starting point exact solutions which can be written in Kerr-Schild form:
\begin{equation} \label{Kerr-SchildMetric} 
    g_{\mu \nu} = \eta_{\mu \nu} + \phi k_{\mu} k_{\nu},
\end{equation}
where $k_{\mu}$ is a null geodesic congruence with respect to the flat background
$\eta_{\mu \nu}$. Such solutions are of interest as they linearize Einstein's equations in terms of the field
$h_{\mu \nu} = \phi k_{\mu} k_{\nu}$.

The single copy field is the Abelian gauge field $A_{\mu} = \phi k_{\mu}$, and $\phi$ is the zeroth copy
scalar field. They satisfy the Maxwell equations and the Klein-Gordon wave equation on the flat background $\eta_{\mu
\nu}$.\footnote{ \label{FlatVCurvedFootnote}  Much of the double copy
literature uses a flat background owing to its origins in the study of amplitudes in flat spacetime. However, the classical double copy can still be performed in curved spacetimes, and the curved equations of motion can be used; for relevant literature, see \cite{Alawadhi:2020jrv, Gurses:2018ckx,Bahjat-Abbas:2017htu,
Carrillo-Gonzalez:2017iyj, Prabhu:2020avf, Adamo:2017nia, Alkac:2021bav, Han:2022mze,Alkac:2021bav,Han:2022ubu}.}

The Weyl formulation \cite{Luna:2018dpt} of
the classical double copy is expressed instead in terms of the Weyl curvature and the field strength tensor.
The index structure of these objects is simpler when written as spinors, in which case the double copy relation is
\begin{equation} \label{WeylSpinorDoubleCopyRelation} 
    \Psi_{ABCD} = \frac{1}{S} f_{(AB} f_{CD)},
\end{equation}
where $\Psi_{ABCD}$ is the Weyl spinor, $f_{AB}$ is the single copy Maxwell spinor and $S$ is the zeroth copy
scalar field. The tensor translation of the preceding equation is \cite{Alawadhi:2019urr, Alawadhi:2020jrv}
\begin{multline} \label{WeylDoublingRelation} 
    W_{\mu \nu \rho \sigma} = \frac{1}{S} \Big( F_{\mu \nu} F_{\rho \sigma} + F_{\mu \rho} F_{\nu \sigma} -
    \frac{6}{D-2} g_{\mu \rho} F_{\nu \alpha} F_{\sigma}^{\phantom{\sigma}\alpha} \\
    + \frac{3}{(D-1)(D-2)}
g_{\mu \rho} g_{\nu \sigma} F_{\alpha \beta} F^{\alpha \beta}\Big) \bigg\vert_{s},
\end{multline}
where ``$ \big\vert_{s}$'' indicates that the expression must be anti-symmetrized pairwise in $\mu, \nu$ and in
$\rho, \sigma$ with unit weight.

While the Kerr-Schild single copy of the higher dimensional Kerr solution was described already in
\cite{Monteiro:2014cda}, we aim to investigate whether instances of a classical Weyl double copy continue to exist
in dimensions greater than four. We propose, as a starting point, a class of well-understood black hole spacetimes
known as the Kerr-NUT-(A)dS family, described in the next section. We will find also a satisfying multi-Kerr-Schild
description of the single copy of Kerr-NUT-(A)dS in arbitrary dimension, which includes as a special case the
Kerr-Schild single copy of higher dimensional Kerr.

\subsection{Higher dimensional black holes}

The Kerr-NUT-(A)dS family of vacuum spacetimes includes all stationary black holes with a spherical horizon topology
and arbitrary mass, angular momentum and NUT parameters.
Following \cite{Frolov:2017kze}, we set $D = 2n + \varepsilon$, where $\varepsilon = 0$
for even $D$ and $\varepsilon = 1$ for odd $D$. The Kerr-NUT-(A)dS metric attains its most
symmetric form in \emph{canonical
coordinates} $(x_{k}, \psi_{j})$,
\begin{equation} \label{KNAmetricInCanonicalCoordinatesGeneralD} 
    ds^{2} = \sum_{k = 1}^{n} \left[ \frac{U_{k}}{X_{k}} dx_{k}^{2} +
    \frac{X_{k}}{U_{k}}\left( \sum_{j = 0}^{n-1} A_{k}^{(j)} d\psi_{j}\right)^{2}\right] -
    \varepsilon \frac{ \mathcal{A}^{(n - 1 + \varepsilon)}}{A^{(n)}}
    \left( \sum_{j = 0}^{n - 1 + \varepsilon} d\psi_{j}\right)^{2}.
\end{equation}
The $\psi_{j}$ coordinates for $i=0$ to $n-1+\epsilon$ are the Killing directions, and are linear combinations of
the azimuthal angles $\phi_{j}$ and time $t$. The $x_{k}$ coordinates correspond to the remaining angles
$\theta_{k}$, except $x_{n} = ir$, which is the Wick-rotated radial coordinate.

Here, the functions $U_{k}$, $X_{k}$, $A_{k}^{(j)}$ and $A^{(j)}$ are
polynomials in $x_{k}$, given by
\begin{align}  \nonumber
        X_{k} &= \frac{-(1 + \lambda x_{k}^{2})}{(-x_{k}^{2})^{\varepsilon}} \prod_{m = 1}^{n - 1 +
        \varepsilon}(a_{m}^{2} - x_{k}^{2}) + 2 M_{k} (-x_{k})^{1 - \varepsilon}, & &U_{k} = \prod_{ \substack{j = 1 \\ j \neq k}}^{n} (x_{j}^{2} - x_{k}^{2}), \\
        A_{k}^{(j)} &= \sum_{ \substack{1 \leq i_{1} < i_{2} < \dots < i_{j} \leq n \\ i_{1}, i_{2},
        \dots i_{j} \neq k}}^{} x_{i_{1}}^{2} x_{i_{2}}^{2} \dots x_{i_{j}}^{2},& &A^{(k)}
        = \sum_{1 \leq i_{1} < i_{2} < \dots < i_{k} \leq n}^{}x_{i_{1}}^{2} x_{i_{2}}^{2} \dots
        x^{2}_{i_{k}},\label{KNAmetricFunctions} 
\end{align}
and $\mathcal{A}_{m}^{(l)}$, $\mathcal{A}^{(l)}$ are polynomials in $a_{m}$
\begin{equation} \label{KNAspinPolynomials} 
    \mathcal{A}_{m}^{(l)} = \sum_{ \substack{1 \leq i_{1} < i_{2} < \dots < i_{l} \leq n-1+\varepsilon \\ i_{1}, i_{2},
        \dots i_{l} \neq m}}^{} a_{i_{1}}^{2} a_{i_{2}}^{2} \dots a_{i_{l}}^{2},
        \qquad \mathcal{A}^{(l)} = \sum_{1 \leq i_{1} < i_{2} < \dots < i_{l} \leq n - 1 + \varepsilon}^{} a_{i_{1}}^{2} \dots a_{i_{l}}^{2}.
\end{equation}
The constants $a_{j}$ are the independent spins of the black hole, and $\lambda$ is related to the cosmological
constant via $\Lambda = \frac{1}{2}(D-2)(D-1) \lambda$. The parameters $M_{k}$ are NUT parameters, except for
$M_{n}$, which is the mass in odd dimensions and $i$ times the mass in even dimensions.

Alternatively, the spacetime metric can be described in canonical coordinates by
exhibiting an orthonormal frame of vectors
\begin{equation} \label{DarbouxFrame} 
    \begin{split}
        \mathbf{e}_{k} &= \sqrt{ \frac{X_{k}}{U_{k}}} \partial_{x_{k}}, \\
        \mathbf{\hat{e}}_{k} &= \sqrt{ \frac{U_{k}}{X_{k}}} \sum_{j = 0}^{n - 1 +
        \varepsilon} (-x_{j}^{2})^{n - 1 - j} \partial_{\psi_{j}}, \\
            \mathbf{\hat{e}}_{0} &= \sqrt{- \mathcal{A}^{(n - 1 + \varepsilon)} A^{(n)}} \partial_{\psi_{n}},
    \end{split}
\end{equation}
which is dual to the orthonormal coframe
\begin{equation} \label{DarbouxCoFrame} 
    \begin{split}
        \mathbf{e}^{k} &= \sqrt{ \frac{U_{k}}{X_{k}}} dx_{k}, \\
        \mathbf{\hat{e}}^{k} &= \sqrt{ \frac{X_{k}}{U_{k}}} \sum_{j = 0}^{n - 1}
        A_{k}^{(j)} d\psi_{j}, \\
        \mathbf{\hat{e}}^{0} &= \sqrt{ \frac{- \mathcal{A}^{(n - 1 + \varepsilon)}}{A^{(n)}}} \sum_{j = 0}^{n} A^{(j)} d\psi_{j}.
    \end{split}
\end{equation}
The negative sign in the square roots will be canceled by the one from squaring $x_{n} = i r$. With the
appropriate coordinate ranges, $\mathbf{\hat{e}}^{n}$ will turn out to be purely imaginary, thus
playing the role of the timelike element of the basis.

Explicitly, the metric in this frame is
\begin{equation} \label{MetricInDarbouxCoFrame} 
    ds^{2} = \sum_{k = 1}^{n} \left( \mathbf{e}^{k} \mathbf{e}^{k} + \mathbf{\hat{e}}^{k}
    \mathbf{\hat{e}}^{k}\right) + \varepsilon \mathbf{\hat{e}}^{0} \mathbf{\hat{e}}^{0}.
\end{equation}
These spacetimes are distinguished in possessing a full \emph{Killing tower}: a non-degenerate set of $n +
\varepsilon$ Killing vectors and $n$ Killing tensors. These geometric objects are generators of explicit and
hidden symmetries, which make geodesic motion in these spacetimes completely integrable.

The Killing tower of Kerr-NUT-(A)dS spacetimes is generated using a single $2$-form, known as the principal
tensor. The existence of a principal tensor is a nontrivial constraint on spacetime geometry, and the
Kerr-NUT-(A)dS metric is (locally) the most general vacuum spacetime which admits a principal tensor
\cite{Houri:2007xz, Frolov:2017kze}. %
In terms of the orthornormal coframe, the principal tensor for the geometry
\eqref{KNAmetricInCanonicalCoordinatesGeneralD} is
\begin{equation} \label{PrincipalTensorInDarbouxCoFrame} 
    \mathbf{h} = \sum_{k = 1}^{n} x_{k} \mathbf{e}^{k} \wedge \mathbf{\hat{e}}^{k}.
\end{equation}

\subsubsection{Transformation to Boyer-Lindquist and Myers-Perry}
While canonical coordinates are well-suited to examining symmetries of the spacetime for
generic spins, mass, cosmological constant and NUT parameters, taking any special limit of
the spin parameters or the cosmological constant in this coordinate system requires a
great deal of care, since the transformation to familiar coordinates is parameter-dependent.

Specifically, the coordinate transformations from $\psi_{j}$ to azimuthal angles and the Killing time, and from
$x_{k}$ to $\theta_{k}$, depend non-trivially on the spin parameters and $\lambda$.
The coordinate transformation from the $n + \varepsilon $ coordinates $\psi_{j}$ to $t$ and the $n-1+\varepsilon$
angles $\phi_{k}$ defines \emph{generalized Boyer-Lindquist coordinates}
\cite{Frolov:2017kze}
\begin{equation} \label{TransformAzimuthalAnglesFromCanonicalToMyersPerry} 
    \begin{split}
        t &= \sum_{k = 0}^{n - 1 + \varepsilon} \mathcal{A}^{(k)} \psi_{k}, \\
        \phi_{j} &= a_{j} \sum_{k = 0}^{n - 1 + \varepsilon} \left( \lambda
        \mathcal{A}_{j}^{(k)} - \mathcal{A}_{j}^{(k-1)}\right) \psi_{k}.
    \end{split}
\end{equation}

We can make a further transformation to
Myers-Perry coordinates $(t, r, \mu_{i}, \phi_{j})$, defining $r$ by $x_{n} = ir$ and the latitude coordinates
$\mu_{i}$ by\footnote{When indexing the latitude coordinates $\mu_{i}$ we follow the conventions of \cite{Chen:2006xh} instead
of \cite{Frolov:2017kze}.} 
\begin{equation} \label{} 
    \mu_{i}^{2} \prod_{ \substack{j
            = 1 \\ j\neq i}}^{n}(a_{i}^{2} - a_{j}^{2})  =  \prod_{k = 1}^{n - 1} (a_{i}^{2} - x_{k}^{2}).
\end{equation}
In even dimensions, the expression above includes the extraneous parameter $a_{n}$. To
obtain the correct transformation, $a_{n}$ is set to $0$ in even dimensions.
The $n$ coordinates $\mu_{i}$ involve only the $n-1$ coordinates $x_{k}$,
and are thus not all independent. They satisfy the constraint
\begin{equation} \label{} 
    \sum_{i = 1}^{n} \mu_{i}^{2} = 1,
\end{equation}
and are best thought of as direction cosines for the independent rotational $2-$planes.

We will now discuss the $D=4$ and $D=5$ cases in detail. They are instructive as to the general form of the
coordinate transformations, and we will later perform new explicit computations in five dimensions, as well as
comparisons with the extensive prior classical double copy literature in four dimensions. In $D = 4$ the
transformation from canonical to Myers-Perry coordinates is (setting $a_{1} = a$) 
\begin{equation} \label{CanonicalToMyersPerryD=4} 
    \begin{gathered}
    t = \psi_{0} + a^{2} \psi_{1},\qquad r = -i x_{2},\qquad \phi = \lambda a \psi_{0} - a \psi_{1}, \\
    \mu_{1}^{2} = \frac{(a^{2} - x_{1}^{2})}{a^{2}}, \qquad \mu_{2}^{2} = \frac{x_{1}^{2}}{a^{2}}.
    \end{gathered}
\end{equation}
The redundant latitude coordinates $\mu_{1}$ and $\mu_{2}$, in terms of the familiar
co-latitude angle $\theta$, are simply $\sin\theta$ and $\cos\theta$. Solving for $x_{1}$ in terms of $\theta$, 
\begin{equation} \label{x1ToThetaD=4} 
    x_{1} = a \cos\theta.
\end{equation}
In the Schwarzschild limit $a \to 0$, $x_{1}$ ceases to be a well-defined coordinate. The coordinate transformation
to $\phi$ is also singular in this limit, and the canonical angles $\psi_{0}, \psi_{1}$ are ill-defined.

In $D=5$, the relation between the Myers-Perry and canonical coordinate systems is 
\begin{align} \label{CanonicalToMyersPerryD=5} 
        t &= \psi_{0} + (a_{1}^{2} + a_{2}^{2}) \psi_{1} + a_{1}^{2} a_{2}^{2}
        \psi_{2}, & r &= -i x_{2}, \nonumber \\
        \frac{\phi_{1}}{a_{1}} &= \lambda \psi_{0} + (\lambda a_{2}^{2} - 1) \psi_{1} -
        a_{2}^{2} \psi_{2}, & \frac{\phi_{2}}{a_{2}} &= \lambda \psi_{0} + (\lambda a_{1}^{2} - 1) \psi_{1} - a_{1}^{2} \psi_{2}, \\
        \mu_{1}^{2} &= \frac{a_{1}^{2} - x_{1}^{2}}{a_{1}^{2} - a_{2}^{2}}, & \mu_{2}^{2} &= \frac{a_{2}^{2} -
        x_{1}^{2}}{a_{2}^{2} - a_{1}^{2}}. \nonumber
\end{align}
Again, setting $\mu_{1}$ and $\mu_{2}$ to $\sin\theta$ and $\cos\theta$, the inverse transformation is
\begin{equation} \label{} 
    x_{1}^{2} = a_{1}^{2} \cos^{2}\theta + a_{2}^{2} \sin^{2}\theta.
\end{equation}
Here, both the spinless limit and the equal-spin limit $a_{2} \to a_{1}$ set $x_{1}$ to a constant (either $0$ or
$a_{1}$) for any value of $\theta$. In either case, $x_{1}$ is an ill-defined coordinate for the spacetime. 

In the following two sections, we describe two properties of Kerr-NUT-(A)dS that make it amenable to a double copy
description: the existence of a multi-Kerr-Schild form of the metric and the algebraic speciality of the Weyl
tensor.

\subsubsection{Multi-Kerr-Schild form} \label{sec:MultiKerrSchild} 

A spacetime whose metric is in the Kerr-Schild form \eqref{Kerr-SchildMetric} has a simple proposal for its
single copy. The vacuum Einstein equations further guarantee that the single copy field is on-shell.
The Kerr-NUT-(A)dS metric \eqref{KNAmetricInCanonicalCoordinatesGeneralD} can be written in a similar way: a multi-Kerr-Schild
form about a flat (or dS/AdS for nonzero $\lambda$) background metric
$\mathring{\mathbf{g}}$ as%
\footnote{The background $\mathbf{\mathring{g}}$ matches
the Kerr-NUT-(A)dS metric when all $M_{k}$ vanish, although it does not take the same form in canonical coordinates.
There is a coordinate system in which it does have the same metric components as the corresponding Kerr-NUT-(A)dS
metric; see \cite{Frolov:2017kze} for details. Such ``Kerr-Schild'' coordinates are convenient when computing the
flat limit of derivatives.}
\begin{equation} \label{MultiKerrSchildMetric} 
    \mathbf{g} = \mathring{\mathbf{g}} + 2 \sum_{k = 1}^{n} \frac{M_{k}x_{k}^{1-\varepsilon}}{U_{k}}
    \boldsymbol{\mu}^{k} \boldsymbol{\mu}^{k},
\end{equation}
where the 1-forms $ \boldsymbol{\mu}^{k}$ are null and geodesic. Note that these 1-forms are unrelated to the
Myers-Perry latitude coordinates $\mu_{i}$ despite their usage of the same symbol. Their explicit form in terms of the orthonormal coframe \eqref{DarbouxCoFrame}
is
\begin{equation} \label{MuDefn} 
        \boldsymbol{\mu}^{k} = \sum_{j = 0}^{n - 1} A_{k}^{(j)} d\psi_{j} + i \frac{U_{k}}{X_{k}} dx_{k} = \sqrt{
        \frac{U_{k}}{X_{k}}} \left(\mathbf{\hat{e}}^{k} + i \mathbf{e}^{k}\right).
\end{equation}
In canonical coordinates, the explicit form of the flat metric $\mathring{\mathbf{g}}$ is%
\footnote{This is essentially eq (4.75) from \cite{Frolov:2017kze}, with a couple of typos corrected (in the most
up-to-date version as of writing) and $\mathbf{\hat{e}}^{0}$ used instead of its scaled version
$\mathbf{\hat{\epsilon}}^{0}$.}
\begin{equation} \label{} 
    \mathring{\mathbf{g}} = \sum_{k = 1}^{n} \frac{X_{k}}{U_{k}} \bigg\vert_{M_{k} = 0} \boldsymbol{\mu}^{k}
    \boldsymbol{\mu}^{k} - i \sum_{k = 1}^{n} \left(\boldsymbol{\mu}^{k} dx_{k} + dx_{k}\
    \boldsymbol{\mu}^{k}\right) +
    \varepsilon\ \mathbf{\hat{e}}^{0} \mathbf{\hat{e}}^{0}.
\end{equation}
Since the single copy and zeroth copy fields (usually, see footnote \ref{FlatVCurvedFootnote}) solve equations of motion on
a flat background, an unambiguous definition of the flat metric is needed. The multi-Kerr-Schild form of the metric
provides an unambiguous flat (or dS/AdS) limit of geometric quantities. In particular, the flat limit of covariant
derivatives ($ \nabla^{(0)}_{\mu}$) is defined as the covariant derivative with respect to $\mathbf{\mathring{g}}$.

In section \ref{sec:GeneralDSingleCopy}, we will use the multi-Kerr-Schild form to propose a single copy gauge
field for the generic Kerr-NUT-(A)dS black hole in arbitrary dimensions.
\subsubsection{Type D}
The generic Kerr-NUT-(A)dS spacetime in general dimension is algebraically special, with a type D Weyl curvature.
We will explain the characteristics of type D spacetimes shortly. For more on algebraic speciality
in higher dimensions see \cite{Coley:2004jv, Coley:2007tp, Pravda:2007ty, Ortaggio:2012jd, Milson:2004jx}.

To express the type D condition, we must first define the boost weight of a tensor in a null-orthonormal basis
$(\mathbf{k}, \mathbf{n}, \mathbf{e}^{i})_{i = 1}^{D-2}$ where $\mathbf{k}$ and $\mathbf{n}$ are null vectors such
that $k_{\mu} n^{\mu} = -1$ and the spacelike vectors $\mathbf{e}_{i}$ are mutually orthogonal as well as
orthogonal to $\mathbf{k}$ and $\mathbf{n}$. Any symmetric tensor $T_{\mu \nu}$, for instance, can be expanded in
this basis as
\begin{equation} \label{} 
T_{\mu \nu} = \underbrace{T_{\mathbf{kk}} k_{\mu} k_{\nu}}_{2} + \underbrace{T_{\mathbf{k}i} k_{(\mu}
e_{\nu)}^{i}}_{1} + \underbrace{T_{\mathbf{kn}} k_{(\mu}n_{\nu)} + T_{ij} e^{i}_{(\mu} e^{j}_{\nu)}}_{0}
+ \underbrace{T_{\mathbf{n}i} n_{(\mu}
e_{\nu)}^{i}}_{-1} + \underbrace{T_{\mathbf{nn}} n_{\mu} n_{\nu}}_{-2},
\end{equation}
where the different terms have been labeled by their boost weight. The boost weight of each term is the power of
$\Lambda$ it scales by under a Lorentz boost in the $(\mathbf{k}, \mathbf{n})$ subspace, which scales $\mathbf{k}$
and $\mathbf{n}$ as
\begin{equation} \label{} 
    \mathbf{k} \to \Lambda \mathbf{k}, \qquad \mathbf{n} \to \Lambda^{-1} \mathbf{n}.
\end{equation}
So, for example, term labeled by the boost weight $2$ transforms as
\begin{equation} \label{} 
    T_{\mathbf{kk}} k_{\mu} k_{\nu} \to T_{\mathbf{kk}} (\Lambda k_{\mu}) (\Lambda k_{\nu}) = \Lambda^{2}\ 
    T_{\mathbf{kk}} k_{\mu} k_{\nu}.
\end{equation}

The Weyl tensor of a spacetime $W_{\mu \nu \rho \sigma}$, too, can be expanded in this basis and its components
classified by boost weight. For a type D spacetime, there always exists a null-orthonormal basis in which its only
nonzero components have boost weight $0$. This can be confirmed by checking that the following conditions hold:
\cite{Hamamoto:2006zf}
\begin{equation} \label{TypeDConditions} 
    \begin{split}
        W^{\mu \nu \rho \sigma} k_{\mu} e_{\nu}^{i} e_{\rho}^{j} e_{\sigma}^{k} &= W^{\mu \nu \rho \sigma}
        n_{\mu}e_{\nu}^{i} e_{\rho}^{j} e_{\sigma}^{k} = 0, \\
        W^{\mu \nu \rho \sigma} k_{\mu} e_{\nu}^{i} k_{\rho}e_{\sigma}^{j} &= W_{\mu \nu \rho \sigma} n_{\mu}
        e_{\nu}^{i} n_{\rho} e_{\sigma}^{j} = 0, \\
        W^{\mu \nu \rho \sigma} k_{\mu} n_{\nu}k_{\rho} e_{\sigma}^{i} &= W^{\mu \nu \rho \sigma} n_{\mu}
        k_{\nu}n_{\rho} e_{\sigma}^{i} = 0,
    \end{split}
\end{equation}
for the basis choice
\begin{equation} \label{}
    \mathbf{k} = \frac{\mathbf{e}_{n} + i \mathbf{\hat{e}}_{n}}{\sqrt{2} }, \qquad \mathbf{n} =
    \frac{\mathbf{e}_{n} - i \mathbf{\hat{e}}_{n}}{\sqrt{2} }.
\end{equation}
The vectors $\mathbf{k}$ and $\mathbf{n}$ are real null vectors, since $\mathbf{\hat{e}}_{n}$ is imaginary once the Wick rotation to Lorentzian signature is
carried out. We will refer to any basis that satisfies \eqref{TypeDConditions} as a Weyl-aligned frame.

\subsection{Spinors in five dimensions} \label{sec:ReviewSpinors5D} 
In four dimensions, the Weyl formulation of the double copy is most compactly expressed using spinor representations
of the Weyl tensor and the Maxwell field strength. This continues to hold in five dimensions, as we demonstrate in
section \ref{sec:FiveDDoubleCopy}. The following is a summary of relevant spinor definitions from
\cite{Monteiro:2018xev}.

In five dimensions the objects which translate between tensor and spinor representations are gamma matrices,
which are a representation of the Dirac algebra
\begin{equation} \label{} 
    \gamma^{(\mu}_{AB} \gamma^{\nu)}_{CD} \Omega^{BC} = g^{\mu \nu} \Omega_{AD},
\end{equation}
where $A = 1, 2, 3, 4$ is a spinor index and $\mu = 0, 1, 2, 3, 4$ is a tensor index. We work with the metric
$g^{\mu \nu}$ in $(- + + + +)$ signature. The representation of gamma matrices used is exactly the one used in
\cite{Monteiro:2018xev}. We use also the antisymmetric bispinor $\Omega_{AB}$ defined in
\cite{Monteiro:2018xev} to provide an antisymmetric product of two arbitrary spinors $\xi^{A}$ and
$\chi^{A}$ via
\begin{equation} \label{} 
    \xi_{A} \chi^{A} = \xi^{A} \Omega_{AB} \chi^{B} = \xi_{A} \Omega^{AB} \chi_{B}.
\end{equation}
We use the following convention for raising and lowering spinor indices
\begin{equation} \label{} 
    \chi_{A} = \chi^{B} \Omega_{BA}, \qquad \chi^{A} = \Omega^{AB} \chi_{B},
\end{equation}
which in turn implies
\begin{equation} \label{} 
    \Omega^{AC} \Omega_{BC} = \mathbbm{1}^{A}_{B}.
\end{equation}

Since our aim will be to work with spinor representations of tensor fields, we will find it of use to work in a
spinor basis that is associated to a (null) orthonormal vector basis.
\subsubsection{Spinor basis and little group} \label{sec:SpinorBasis5D} 
The spinor basis we work in is defined in terms of a null frame of vectors comprising of real null vectors
$k^{\mu}$ and $n^{\mu}$ satisfying $k^{\mu} n_{\mu} = -1$, a complex null vector $m^{\mu}$ and its complex
conjugate $\overline{m}^{\mu}$ which obey $m^{\mu} \overline{m}_{\mu} = 1$ and a spacelike unit vector
$e_{0}^{\mu}$. 

In other words, the inverse metric in terms of the null frame is
\begin{equation} \label{NullFrameMetric} 
    g^{\mu \nu} = -k^{\mu} n^{\nu} - n^{\mu} k^{\nu} + m^{\mu} \overline{m}^{\nu} + \overline{m}^{\mu}
    m^{\nu} + e_{0}^{\mu} e_{0}^{\nu}.
\end{equation}

The spinor basis $(k^{A}_{a}, n^{A}_{a})$ associated to the null frame is defined by
\begin{equation} \label{SpinorBasisDefn} 
        k_{\mu} \gamma^{\mu}_{AB} k^{A}_{a} = 0, \qquad n_{\mu} \gamma^{\mu}_{AB} n^{A}_{a} = 0,
\end{equation}
where $a = 1,2$, and
\begin{equation} \label{SpinorBasisNormalization} 
    k_{a}^{A} \Omega_{AB} n^{B}_{b} = \epsilon_{ab}.
\end{equation}
The convention used for the two-dimensional Levi-Civita symbol $\epsilon_{ab}$ is $\epsilon^{12} = -\epsilon_{12}
= 1$. It is used to lower and raise lowercase latin indices in the same way $\Omega_{AB}$ is used to lower and
raise spinor indices.

This definition does not uniquely fix a basis relative to a particular null frame $(k^{\mu}, n^{\mu}, m^{\mu}, 
\overline{m}^{\mu}, e_{0}^{\mu})$. Any two bases which satisfy \eqref{SpinorBasisDefn} and
\eqref{SpinorBasisNormalization} are related by an $SU(2)$ action on lowercase latin indices.
This $SU(2)$ is referred to as the little group.

Contractions of any spinor with the spinor basis will carry $SU(2)$ indices and
thus transform under the little group. We briefly look at contractions of the Weyl spinor and the resulting little
group objects in the next subsection.
\subsubsection{Weyl little group objects} \label{sec:WeylSpinorLittleGroupObjects} 
As hinted at in the beginning of this section, we will find it convenient to work with the spinor representation of
the Weyl tensor. Owing to the symmetries of the Weyl tensor $W_{\mu \nu \rho \lambda}$, all its independent components can be packaged into
a four-index spinor $\Psi_{ABCD}$ which is symmetric under any permutation of its indices. This Weyl spinor is constructed with
the aid of the spinor Lorentz generators
\begin{equation} \label{SpinorLorentzGeneratorsDefn} 
    \sigma^{\mu \nu}_{AB} = \gamma^{[\mu}_{AC} \gamma^{\nu]}_{DB} \Omega^{CD},
\end{equation}
in the following way
\begin{equation} \label{WeylSpinorDefn} 
    \Psi_{ABCD} = W_{\mu \nu \rho \lambda} \sigma^{\mu \nu}_{AB} \sigma^{\rho \lambda}_{CD}.
\end{equation}
Following \cite{Monteiro:2018xev} the Weyl spinor can be expanded in our chosen frame-associated spinor basis
\begin{multline} \label{WeylLittleGroupObjects} 
    \Psi_{ABCD} = \Psi^{(0)}_{abcd}\ k^{a}_{(A} k^{b}_{B} k^{c}_{C} k^{d}_{D)} + 4
    \Psi^{(1)}_{abcd}\ k^{a}_{(A} k^{b}_{B} k^{c}_{C} n^{d}_{D)} + 6 \Psi^{(2)}_{abcd}\
    k^{a}_{(A} k^{b}_{B} n^{c}_{C} n^{d}_{D)} \\
    + 4 \Psi^{(3)}_{abcd}\ k^{a}_{(A} n^{b}_{B}
    n^{c}_{C} k^{d}_{D)} + \Psi^{(4)}_{abcd}\ n^{a}_{(A} n^{b}_{B} n^{c}_{C} n^{d}_{D)}.
\end{multline}
The components $\Psi^{(i)}_{abcd}$ carry only $SU(2)$ indices ${a,b,c,d}$ and are dubbed ``little group objects''. Based on their index symmetry, each component can be broken into their
respective irreducible representations (irreps) under $SU(2)$.

Since the spacetimes we will consider are type D, we note that the only nonzero little
group object is $\Psi^{(2)}_{abcd}$ if $k^{\mu}$ and $n^{\mu}$ are chosen such that the type D conditions
\eqref{TypeDConditions} hold; see Appendix \ref{SpinorConventions}. The irreps contained within
$\Psi^{(2)}_{abcd}$ are the fully symmetric objects $\psi^{(2)_{abcd}}$ and $\chi^{(2)}_{ab}$, and the scalar
$\Psi^{(2)}_{\mathrm{tr}}$. In terms of these irreps, $\Psi^{(2)}_{abcd}$ is
\begin{multline} \label{LittleGroup2Irreps} 
    \Psi^{(2)}_{abcd} = \psi^{(2)}_{(abcd)} - \frac{1}{4} \left( \epsilon_{ac}
    \chi^{(2)}_{(bd)} + \epsilon_{ad} \chi^{(2)}_{(bc)} + \epsilon_{bc} \chi^{(2)}_{(ad)}
+ \epsilon_{bd} \chi^{(2)}_{(ac)}\right) + \\
\frac{1}{6} \left( \epsilon_{ac}\epsilon_{bd} +
\epsilon_{ad} \epsilon_{bc}\right) \Psi^{(2)}_{\mathrm{tr}}.
\end{multline}
Similarly to the expansion of $\Psi^{(2)}_{ABCD}$, in general the little group objects
$\psi^{(2)}_{abcd}$ and $\chi^{(2)}_{ab}$ can be expanded in a little group basis $(\alpha_{a}, \beta_{b})$ as
\begin{equation} \label{IrrepExpansion} 
    \begin{split}
        \psi^{(2)}_{abcd} = &\ \psi_{0}^{(2)} \alpha_{a} \alpha_{b} \alpha_{c} \alpha_{d} + 4
    \psi_{1}^{(2)} \alpha_{(a} \alpha_{d} \alpha_{c} \beta_{d)} + 6\psi^{(2)}_{2} \alpha_{(a} \alpha_{b} \beta_{c}
    \beta_{d)} + 4 \psi^{(2)}_{3} \alpha_{(a} \beta_{b} \beta_{c} \beta_{d)}\\
                            & + \psi^{(2)}_{4} \beta_{a} \beta_{b} \beta_{c} \beta_{d}, \\
        \chi^{(2)}_{ab} = &\ \chi_{0}^{(2)} \alpha_{a} \alpha_{b} + 2 \chi_{1}^{(2)} \alpha_{(a}
    \beta_{b)} + \chi_{2}^{(2)} \beta_{a} \beta_{b}.
    \end{split}
\end{equation}
Analogous to how algebraic speciality indicates the existence of a spinor basis in terms of which various little
group objects $\Psi^{(i)}_{abcd}$ may vanish, within each algebraically special class we can make a finer
classification based on the existence of little group bases $(\alpha_{a}, \beta_{a})$ for which little group
irreps take special forms. For instance, for certain spacetimes within the type D class it may be possible to
choose a basis $(\alpha_{a}, \beta_{a})$ such that the only nonzero components of $\psi^{(2)}_{abcd}$
and $\chi^{(2)}_{ab}$ are $\psi_{2}^{(2)}$ and $\chi^{(2)}_{1}$. Precisely such a fine-grained notion of algebraic
speciality, within the coarse type D class, will be of use in proposing a single copy of Kerr-NUT-(A)dS in five
dimensions. That these finer classifications could be important was anticipated already in \cite{Monteiro:2018xev}.

\section{Single copy of five dimensional black holes} \label{sec:FiveDDoubleCopy} 
The general Kerr-NUT-(A)dS spacetime in five dimensions includes all stationary black holes with
spherical horizon topology and arbitrary values of mass, angular momentum, NUT charge and
cosmological constant.
In this section we compute the Weyl spinor for these spacetimes and show that it can be written as the double copy
of a Maxwell field. If we interpret the field as living on the curved
spacetime instead of on the flat background (as is otherwise usual for single copies, see \cite{Alawadhi:2020jrv, Bahjat-Abbas:2017htu,
Carrillo-Gonzalez:2017iyj, Prabhu:2020avf, Adamo:2017nia, Alkac:2021bav, Han:2022mze} for examples
where the field lives on a curved background), it turns out to belong to a
previously known class of Maxwell fields on Kerr-NUT-(A)dS which are aligned along the principal tensor of the
spacetime \cite{Krtous:2007xg}. 

In fact, many previously known examples of the double copy map --- in particular the Schwarzschild-Tangherlini and
Myers-Perry black holes in general dimension and the $D = 4$ Taub-NUT solution --- can be described as aligned
fields double-copying to Kerr-NUT-(A)dS solutions.

\subsection{Curvature tensors and Weyl spinor}

We now exhibit the Weyl curvature of the five-dimensional
Kerr-NUT-(A)dS black hole, both in terms of the Weyl tensor and the Weyl spinor. We 
show that for this class of spacetimes, the Weyl spinor has a well-defined square root,
which is algebraically necessary for a double-copy description that involves a map to
exactly one single copy field.

Beginning from the components of the Riemann and Ricci curvature tensors for the general
Kerr-NUT-(A)dS spacetime computed in \cite{Hamamoto:2006zf}, we can compute the Weyl
tensor $ W_{\mu \nu \rho \sigma}$ using 
\begin{multline} \label{WeylTensorFromRiemann} 
    W_{\mu \nu \rho \sigma} = R_{\mu \nu \rho \sigma} - \frac{1}{D-2}(R_{\mu \rho} g_{\nu
    \sigma} - R_{\nu \rho} g_{\mu \sigma} + R_{\nu \sigma} g_{\mu \rho} - R_{\mu\sigma}
    g_{\nu \rho})  \\ +\frac{R}{(D-1)(D-2)} (g_{\mu \rho} g_{\nu \sigma} - g_{\nu \rho}
    g_{\mu \sigma}).
\end{multline}
In the Darboux basis \eqref{DarbouxCoFrame} the Weyl tensor is
\begin{equation} \label{WeylInFormNotation5D} 
    \begin{split}
        \mathbf{W} = &\sum_{i=1}^{n} \left(- \frac{1}{2} \partial_{i}^{2} \hat{Q} -
                \frac{2r_{i}}{3} + \frac{\sum_{k}^{}r_{k}}{6}\right)\left(\mathbf{e}^{i}
        \wedge \mathbf{\hat{e}}^{i}\right)^{2} \\
                        &+ \sum_{i=1}^{n} \left( - \frac{1}{2x_i} \partial_i \hat{Q} - \frac{r_0
                        r_i}{3} + \frac{r_0 + 2\sum_k r_k}{12}\right)
                        \left[\left( \mathbf{e}^{i} \wedge \mathbf{\hat{e}}^{0}
                            \right)^{2} + \left( \mathbf{\hat{e}}^{i} \wedge
                    \mathbf{\hat{e}}^{0} \right)^{2} \right] \\
          &+  \sum_{ \substack{1 \leq i < j \leq n}}^{} \left(\frac{x_{j} \partial_{j} \hat{Q} - x_{i}
              \partial_{i}\hat{Q} }{2(x_{i}^{2} - x_{j}^{2})} - \frac{r_{i} + r_{j}}{3} +
              \frac{\sum_{k}^{}r_{k}}{6} \right)\begin{bmatrix}
                             \left(\mathbf{e}^{i} \wedge \mathbf{e}^{j}\right)^{2} +
                             \left(\mathbf{e}^{i} \wedge \hat{\mathbf{e}}^{j}\right)^{2}
                             \\
                             + \left(\hat{\mathbf{e}}^{i} \wedge
                             \hat{\mathbf{e}}^{j}\right)^{2} + \left(\hat{\mathbf{e}}^{i}
                         \wedge \mathbf{e}^{j}\right)^{2}
                         \end{bmatrix} \\
              &+  \sum_{ \substack{1 \leq i < j \leq 2}}^{} \frac{x_{i}
    \partial_{j} \hat{Q} - x_{j} \partial_{i} \hat{Q}}{x_{i}^{2} - x_{j}^2}
    \begin{bmatrix}
                            \left(\mathbf{e}^{i} \wedge \mathbf{e}^{j}\right) \odot
                            \left(\hat{\mathbf{e}}^{i} \wedge \hat{\mathbf{e}}^{j}\right)
                            \\
                            + \left(\mathbf{e}^{i} \wedge \hat{\mathbf{e}}^{j}\right)
                            \odot \left(\mathbf{e}^{j} \wedge \hat{\mathbf{e}}^{i}\right)
                            \\
                         + 2 \left(\mathbf{e}^{i} \wedge \hat{\mathbf{e}}^{i}\right) \odot
                         \left(\mathbf{e}^{j} \wedge \hat{\mathbf{e}}^{j}\right)
                     \end{bmatrix}.
    \end{split}
\end{equation}
Here, the $ \odot$ represents the symmetrized tensor product, in this case of $2$-forms. The square of a $2$-form,
such as that of $\mathbf{e}^{i} \wedge \mathbf{\hat{e}}^{i}$ in the first line, is shorthand for a tensor product
of two copies of the $2$-form. The symbol $\wedge$ denotes the standard exterior product of differential
forms. The $ r_{i}$ are components of the Ricci tensor
\begin{equation} \label{} 
    \mathbf{Ric} = \sum_{i = 1}^{n} r_{i} \left[ \mathbf{e}^{i} \mathbf{e}^{i} +
    \mathbf{\hat{e}}^{i} \mathbf{\hat{e}}^{i}\right] + r_{0} \mathbf{\hat{e}}^{0}
    \mathbf{\hat{e}}^{0},
\end{equation}
given by
\begin{equation} \label{} 
    r_{i} = - \frac{1}{2} \partial_{i}^{2} \hat{Q} + \sum_{ \substack{j = 1 \\ j\neq i}}^{n}
    \frac{x_{j} \partial_{j} \hat{Q} - x_{i} \partial_{i} \hat{Q}}{x_{i}^{2} - x_{j}^{2}}, 
    \qquad
    r_{0} =
    \sum_{i = 1}^{n} - \frac{1}{x_{i}} \partial_{i}\hat{Q}.
\end{equation}
$ \hat{Q}$ is defined in terms of metric functions \eqref{KNAmetricFunctions} as\footnote{Our definition of
    $\hat{Q}$ is consistent with the definitions of the similarly named functions $Q_{T}$ in even dimensions and
$\hat{Q}_{T}$ in odd dimensions in \cite{Hamamoto:2006zf}.}
\begin{equation} \label{} 
    \hat{Q} = \sum_{k = 1}^{n} \frac{X_{k}}{U_{k}} - \varepsilon \frac{ \prod_{j = 1}^{n - 1 +
    \varepsilon} a_{j}^{2}}{ \prod_{i  = 1}^{n} x_{i}^{2}}.
\end{equation}
The Weyl tensor \eqref{WeylInFormNotation5D}  for Kerr-NUT-(A)dS exhibits extra algebraic symmetries beyond those required of all Weyl curvature tensors. For instance, swapping
$\mathbf{e}^{i}$ and $\mathbf{\hat{e}}^{i}$ leaves the Weyl tensor unchanged.

We show in the next section that the $D = 5$ Weyl spinor exhibits these
extra symmetries in a particularly simple form. A straightforward argument then shows that a
well-defined square root $f_{AB}$ in the sense of \eqref{WeylSpinorDoubleCopyRelation} exists, which allows for
its interpretation as a single copy field. 

\subsection{Constructing the single copy}

Since the spinor definitions used are dimension-dependent, we now specialize to the five dimensional case
throughout this section. We follow the conventions of \cite{Monteiro:2018xev}, and provide a brief review in
section \ref{sec:ReviewSpinors5D} with a few further details in Appendix \ref{SpinorConventions}.

As mentioned in section \ref{sec:WeylSpinorLittleGroupObjects} and shown explicitly in Appendix
\ref{SpinorConventions}, for a type D spacetime in a Weyl-aligned frame (i.e. a frame that satisfies the type D
conditions \eqref{TypeDConditions}) the only nonzero little group object is $\Psi^{(2)}$
\eqref{WeylLittleGroupObjects}. Thus, in the $(k^{a}_{A}, n^{a}_{A})$ basis the Weyl spinor is 
\begin{equation} \label{TypeDWeylSpinorInSpinorBasis} 
    \Psi_{ABCD} = 6 \Psi^{(2)}_{abcd} k^{a}_{(A} k^{b}_{B} n^{c}_{C} n^{d}_{D)}.
\end{equation}

The irreps \eqref{LittleGroup2Irreps} of $\Psi^{(2)}_{ABCD}$ are
\begin{equation} \label{} 
    \begin{split}
    \psi^{(2)}_{abcd} &= \frac{128(M-N)x_{1}^{2}}{(r^{2} + x_{1}^{2})^{3}} 
    \alpha_{(a}\alpha_{b}\beta_{c}\beta_{d)}, \\
    \chi^{(2)}_{ab} &= -i \frac{128(M-N) r x_{1}}{(r^{2} + x_{1}^{2})^{3}} \alpha_{(a}
    \beta_{b)}, \\
    \Psi^{(2)}_{\mathrm{tr}} &= \frac{32(M-N)(x_{1}^{2}-3r^{2})}{(r^{2} + x_{1}^{2})^{3}},
    \end{split}
\end{equation}
where
\begin{equation} \label{} 
    \alpha = 
    \begin{bmatrix}
        1 & 0
    \end{bmatrix}, \qquad
    \beta = 
    \begin{bmatrix}
        0 & 1
    \end{bmatrix}.
\end{equation}
In the $(\alpha_{a}, \beta_{a})$ basis, the symmetric four-index object $\psi^{(2)}_{(abcd)}$ has a component only along $\alpha_{(a}
\alpha_{b} \beta_{c} \beta_{d)}$ and $\chi^{(2)}_{(ab)}$ is only along $\alpha_{(a} \beta_{b)}$.
In terms of the
general expansion of $\psi^{(2)}$ and $\chi^{(2)}$ \eqref{IrrepExpansion}, the only nonzero components are $\psi^{(2)}_{2}$ and
$\chi^{(2)}_{1}$. Additionally, $\psi^{(2)}$ is the ``square'' of
$\chi^{(2)}$: $\psi^{(2)}_{abcd} \sim \chi^{(2)}_{(ab} \chi^{(2)}_{cd)}$. 

The algebraic structure of the these irreps suggests the single copy ansatz $f_{AB} =
f^{(1)}_{ab} k^{a}_{(A} n^{b}_{B)}$ where $f^{(1)}_{ab} = f^{(1)}_{1} \alpha_{a} \beta_{b}
+ f^{(1)}_{2} \beta_{a} \alpha_{b}$. The Weyl polynomial $\mathcal{W}$ provides a straightforward method for computing the `square root' and thus finding $f_{AB}$.  It is defined using a formal parameter $\xi^A$:
\begin{equation} \label{WeylPolynomialDefinition} 
    \mathcal{W} = \Psi_{ABCD} \xi^{A} \xi^{B} \xi^{C} \xi^{D}.
\end{equation}
Since the Weyl spinor $\Psi_{ABCD}$ is symmetric in all indices, its contraction with four
copies of $\xi^{A}$ is an equivalent object. In terms of the variables
\begin{equation} \label{} 
    K^{a} = k^{a}_{A} \xi^{A}, \qquad N^{a} = n^{a}_{A} \xi^{A},
\end{equation}
the Weyl polynomial is
\begin{equation} \label{} 
    \mathcal{W} =  \frac{192(N-M)}{(r^{2} + x_{1}^{2})^{3}} \left[ (r + i x_{1}) K^{1} N^{2}
    - (r - i x_{1}) K^{2} N^{1}\right]^{2}.
\end{equation}
Its form suggests the single copy 
\begin{equation} \label{} 
    f_{AB} = \sqrt{ \frac{192(N-M) S}{(r^{2} + x_{1}^{2})^{3}}} \left[ (r + i
    x_{1}) k^{1}_{(A} n^{2}_{B)} - (r - i x_{1}) k^{2}_{(A} n^{1}_{B)}\right].
\end{equation}
Here, $S$ is the scalar factor that appears in the double copy relation
\eqref{WeylSpinorDoubleCopyRelation}, which we fix%
\footnote{Our procedure to fix $S$ closely follows the treatment for type N spacetimes in \cite{Godazgar:2020zbv},
which also finds a set of first order differential equations for $S$ that are integrable. This approach differs
slightly from the one used for type D spacetimes in \cite{Luna:2018dpt}, which relies on prior work showing how the
double, single and zeroth copies can all be written in terms of a single Killing spinor that always exists in type D.}
by requiring that $F_{\mu \nu} =
\frac{1}{8} f_{AB} \sigma^{AB}_{\mu \nu}$ satisfy Maxwell's equations on the
Kerr-NUT-(A)dS background. This single copy structure of the Weyl tensor was found already in
\cite{Didenko:2011ir} for the case with no NUT parameter.  

Away from sources, either the Maxwell equations $ \nabla^{\mu} F_{\mu \nu} = 0$ or their dual $ \nabla_{[\mu}
F_{\nu \rho]} = 0$ are equivalent to the two independent first order
equations 
\begin{align} \label{} 
    r \sqrt{S} + \left(r^{2} + x_{1}^{2}\right) \frac{\partial}{ \partial r} \sqrt{S} &=
    0, \\
    x_{1} \sqrt{S}  + \left(r^{2} + x_{1}^{2}\right) \frac{\partial}{ \partial x_{1}}
    \sqrt{S} &= 0.
\end{align}
These equations have the unique solution (up to an overall constant)
\begin{equation} \label{ScalarFromMaxwellEquations} 
    S = \frac{q^{2}}{192(N - M)(r^{2} + x_{1}^{2})}.
\end{equation}
With this solution for $S$, the field strength $F_{\mu \nu}$ becomes
 \begin{equation} \label{} 
     F_{\mu \nu} = \frac{q}{2(r^{2} + x_{1}^{2})^{2}} \left(x_{1} \mathbf{e}^{1}_{[\mu}
     \mathbf{\hat{e}}^{1}_{\nu]} - i r \mathbf{e}^{2}_{[\mu}
 \mathbf{\hat{e}}^{2}_{\nu]}\right).
 \end{equation}
This field strength belongs to a class of field strengths known in the
Kerr-NUT-(A)dS literature as \emph{aligned fields}. These are on-shell electromagnetic
fields that, in the orthonormal coframe, have the same structure as the principal tensor
\eqref{PrincipalTensorInDarbouxCoFrame}. That is, they have the expansion \cite{Krtous:2007xg}
\begin{equation} \label{AlignedFieldExpansion} 
    \mathbf{F} = \sum_{k = 1}^{n} f_{k} \mathbf{e}^{k} \wedge \mathbf{\hat{e}}^{k},
\end{equation}
where
\begin{equation} \label{AlignedFieldComponents} 
    \begin{split}
        f_{k} &= \frac{e_{k}}{U_{k}} + 2 x_{k} \sum_{ \substack{j = 1 \\ j\neq k}}^{n} \frac{1}{U_{j}} \frac{e_{j} x_{j}
        - e_{k}x_{k}}{x_{j}^{2} - x_{k}^{2}}; \qquad \text{when }D = 2n \\
            f_{k} &= 2 x_{k} \sum_{j = 1}^{n} \frac{1}{U_{j}} \frac{e_{j} - e_{k}}{x_{j}^{2} - x_{k}^{2}};
            \qquad \text{when }D = 2n + 1.
    \end{split}
\end{equation}
Aligned fields in
general depend on $n$ parameters $e_{k}$; the constant $q$ here corresponds to $2(e_{1} -
e_{2})$. 

Also, we find that the scalar $S$ satisfies the source-free wave equation only in the flat limit (defined in terms of the multi-Kerr-Schild expansion
\eqref{MultiKerrSchildMetric}) 
 \begin{equation} \label{} 
     \square^{(0)} S = 0,
 \end{equation}
 where $ \square^{(0)}$ signifies the Laplacian on the flat
 background.

It is curious that the field strength satisfies Maxwell's equations on the full (curved) background as
well as on the flat background, while the scalar $S$
satisfies the source-free wave equation only when the flat limit is taken. Rather than investigate this issue further, we now instead compare our proposal of the single copy with known examples.

\subsection{Single copy from Weyl prescription consistent with Kerr-Schild and with Weyl
Doubling}
The first work \cite{Monteiro:2014cda}%
\footnote{See especially section 4.2.}
 on the classical double copy described the single copy map for an
arbitrary spin, asymptotically flat black hole with no NUT parameters in general dimension. Translating their proposal for the single copy gauge field to
Myers-Perry coordinates, we find
\begin{equation} \label{} 
    \mathbf{A} = \frac{2M}{U} \left( dt + \frac{U}{V - 2M} dr + a_{1} \mu_{1}^{2}
    d\phi_{1} + a_{2} \mu_{2}^{2} d\phi_{2}\right),
\end{equation}
where
\begin{equation} \label{} 
    V = \frac{1}{r^{1 + \varepsilon}} \prod_{j = 1}^{n - 1 + \varepsilon} (r^{2} +
    a_{j}^{2}), \qquad U = V\left(1 - \sum_{k  = 1}^{n - 1 + \varepsilon}
    \frac{a_{k}^{2}\mu_{k}^{2}}{r^{2} + a_{k}^{2}}\right).
\end{equation}
Transforming further to canonical coordinates, we have
\begin{equation} \label{} 
    \mathbf{A} = \frac{2M}{U_{2}} \left( d\psi_{0} + x_{1}^{2} d\psi_{1} \right) + i
    \frac{2M}{X_{2}} dx_{2}.
\end{equation}
The second term here is pure gauge, since $X_{2}$ is a function of
$x_{2}$ only.%
\footnote{The Kerr-Schild prescription for the double copy is sensitive to gauge choice on the
single copy side. The original single copy for Schwarzschild, for instance, was in an unusual gauge
\cite{Monteiro:2014cda}. \cite{Godazgar:2021iae} discusses this gauge issue in the Weyl double copy context, while \cite{Bahjat-Abbas:2020cyb} relates monopoles in the double copy to singular gauge transformations.} %
The first term matches the aligned
field gauge potential found in \cite{Krtous:2007xg} when $e_{1} = 0,\ e_{2} = 2M$. Thus, the single copy for $D = 5$ Myers-Perry proposed in
\cite{Monteiro:2014cda} is an aligned field.

In addition to matching with the previously known Kerr-Schild single copy, the tensor-language formalism proposed
in \cite{Alawadhi:2020jrv} also agrees with our single copy prescription found using spinors. We take the generic aligned field strength in $D = 5$, 
\begin{equation} \label{} 
    \mathbf{F} = \frac{2x_{1}}{(x_{1}^{2} + r^{2})^{2}}(e_{1} - e_{2}) \mathbf{e}^{1} \wedge \mathbf{\hat{e}}^{1} +
    \frac{2ir}{(x_{1}^{2} + r^{2})^{2}} (e_{2} - e_{1}) \mathbf{e}^{2} \wedge \mathbf{\hat{e}}^{2},
\end{equation}
and plug it into the Weyl doubling relation \eqref{WeylDoublingRelation}. We find that the relation is
satisfied for the scalar choice
\begin{equation} \label{} 
    S = \frac{(e_{1} - e_{2})^{2}}{2(M_{1} - M_{2})(x_{1}^{2} + r^{2})}.
\end{equation}

Our observation that the single copy for Kerr-NUT-(A)dS black holes is an aligned field is not restricted to $D  =
5$. Previously known single copies for $D = 4$ Schwarzschild, Kerr and Taub-NUT spacetimes\cite{Monteiro:2014cda,
Luna:2018dpt, Luna:2015paa} (all of which are in the Kerr-NUT-(A)dS class) are also aligned fields, as we
now briefly demonstrate.

The single copy of the Kerr solution is the gauge field \cite{Luna:2018dpt}%
\footnote{\cite{Luna:2018dpt} claims a slightly different gauge
    field since it sets $M = 0$ in order to take the flat background limit. Even though the $dr$ term is pure
    gauge anyway, it is useful to transform to coordinates $(\mathring{t}, r, \theta,
    \mathring{\phi})$, whose relation to the Kerr-Schild coordinates $(x_{k},
    \mathring{\psi}_{j})$ defined in equation (4.77) of
    \cite{Frolov:2017kze} parallels the transformation to Boyer-Lindquist coordinates from canonical
    coordinates \eqref{TransformAzimuthalAnglesFromCanonicalToMyersPerry}. Then, we can 
   view $\mathbf{A}$ as living on the flat background with metric
$\mathbf{\mathring{g}}$.
Written in these coordinates, the gauge potential takes precisely the form in \cite{Luna:2018dpt} with no parameter
limits post hoc.}
\begin{equation} \label{} 
    \mathbf{A} = \frac{Qr}{r^{2} + x_{1}^{2}} \left(dt + \frac{r^{2}+x_{1}^{2}}{r^{2} + a^{2} - 2Mr}\ dr + a\left( 1
    - \frac{x_{1}^{2}}{a^{2}}\right) d\phi\right).
\end{equation}
Transforming to canonical coordinates, 
\begin{equation} \label{} 
    \mathbf{A} = \frac{Qr}{r^{2} + x_{1}^{2}} \left(d\psi_{0} + x_{1}^{2} d\psi_{1}\right) +
    \frac{Qx_{2}}{X_{2}}
    dx_{2}.
\end{equation}
As in $D = 5$, this is the gauge potential for an aligned field with $e_{1} = 0, e_{2} = - i Q$ up to the second
(pure gauge) term.

The single copy field strength of the (non-rotating) Taub-NUT solution is (up to an overall constant)
\cite{Luna:2015paa}
\begin{equation} \label{} 
    \mathbf{F} = \frac{M}{r^{2}} dt \wedge dr + N \sin \theta d\theta \wedge d\phi,
\end{equation}
where $M$ is the mass and $N$ is the NUT charge. The easiest way to see that this is an aligned field is to
convert the generic aligned field to Myers-Perry coordinates and set the spin $a$ and the
cosmological constant $\lambda$ to $0$. A term-by-term comparison then shows that the aligned field charges
$e_{1}$ and $e_{2}$ must be set to $N$ and $-iM$.

In the next section we show that aligned fields continue to serve as single copies for Kerr-NUT-(A)dS spacetimes in
arbitrary dimensions, with aligned field charges set by the NUT charges and the mass.

\section{Towards a single copy of general dimension Kerr-NUT-(A)dS} \label{sec:GeneralDSingleCopy} 
We have shown that the single copy of Kerr-NUT-(A)dS black holes in $D = 5$ is an aligned electromagnetic
field. We also demonstrated that the various $D = 4$ single copy results for black holes in the Kerr-NUT-(A)dS
class are aligned fields too. These observations lead us to ask: is the single copy of a Kerr-NUT-(A)dS black
hole always an aligned field?

One way to propose a single copy is to use the Kerr-Schild prescription. We start with the multi-Kerr-Schild form of the Kerr-NUT-(A)dS metric
\eqref{MultiKerrSchildMetric} and propose the single copy gauge field
\begin{equation} \label{} 
    \mathbf{A} = \sum_{k = 1}^{n} \frac{2Q_{k}x_{k}^{1-\varepsilon}}{U_{k}} \boldsymbol{\mu}^{k},
\end{equation}
where $ \boldsymbol{\mu}^{k}$ are the null geodesic $1$-forms \eqref{MuDefn} and $Q_{k}$ are constant parameters which satisfy
$Q_k=Q M_k,$ where $Q$ is an arbitrary proportionality constant.

Expanding each $ \boldsymbol{\mu}^{k}$ in canonical
coordinates,
\begin{equation} \label{} 
    \begin{split}
        \mathbf{A} &= \sum_{k = 1}^{n} \frac{2Q_{k}x_{k}^{1-\varepsilon}}{U_{k}} \left( \sum_{j = 0}^{n-1}
        A_{k}^{(j)} d\psi_{j} + i \frac{U_{k}}{X_{k}} dx_{k}\right) \\
                   &= \sum_{k = 1}^{n} 2Q_{k}x_{k}^{1-\varepsilon} \left( \sum_{j = 0}^{n-1}
                   \frac{A_{k}^{(j)}}{U_{k}} d\psi_{j} + i \frac{dx_{k}}{X_{k}}\right).
    \end{split}
\end{equation}
The second term inside the parentheses is gauge trivial since $X_{k}$ is a function of $x_{k}$ only. Thus, this
field is gauge equivalent to
\begin{equation} \label{} 
    \mathbf{A} = \sum_{k = 1}^{n} \frac{2Q_{k}x_{k}^{1-\varepsilon}}{U_{k}} \sum_{j = 0}^{n-1} A_{k}^{(j)}
    d\psi_{j},
\end{equation}
which is the gauge potential of an aligned field \cite{Krtous:2007xg}. Analogous to how the Kerr-Schild double
copy was extended to allow for the single copy of the $D=4$ Taub-NUT spacetime in \cite{Luna:2015paa}, this
gauge field is the multi-Kerr-Schild prescription for the single copy of Kerr-NUT-(A)dS in any dimension. 

In the
case of vanishing NUT charges, the corresponding prescription would be $Q_{k} = 0$ for $k = 1$ to $n - 1$; only $Q_n\propto M$ remains. The
single copy gauge field is then exactly the same as the one proposed as the single copy for higher
dimensional Kerr spacetimes in \cite{Monteiro:2014cda}.

Instead of constructing higher dimensional spinor versions of the Weyl tensor and the
aligned field, we now use our proposal to test a straightforward extension of the Weyl doubling prescription
\eqref{WeylDoublingRelation}. However, plugging in the Weyl tensor components
\eqref{WeylInFormNotation5D} and the components of the aligned field from \eqref{AlignedFieldExpansion}
and \eqref{AlignedFieldComponents}, there is no scalar that satisfies \eqref{WeylDoublingRelation} for any value of
aligned field parameters $Q_{k}$ given arbitrary mass and NUT charges.

Even setting all NUT charges to zero (where our proposal matches \cite{Monteiro:2014cda}), there is no zeroth copy
scalar which satisfies the Weyl doubling prescription \eqref{WeylDoublingRelation}. We expect that the Weyl doubling formula may need additional terms in $D > 5$ in order to describe the
Weyl double copy structure of Kerr-NUT-(A)dS.
\cite{Alawadhi:2020jrv}.

\section{Discussion}\label{sec:discussion}

We have shown that the single copies of the Kerr-NUT-(A)dS spacetimes map onto the `aligned fields' in the same
spacetimes, with charges proportional to the mass and NUT charges.  In five dimensions, we further showed that this
aligned field single copy matches with the Weyl double copy in spinor language and in
tensor language via the Weyl doubling procedure.  

Importantly, this five-dimensional single copy which follows the Weyl doubling formula has arbitrary spins (as well as a possible nonzero NUT charge and cosmological constant), in contrast to the comment in \cite{Alawadhi:2020jrv} that only the singly-rotating solution was possible.  However, we do note that our zeroth-copy scalar only satisfies the flat-space Klein-Gordon equation (the aligned field itself satisfies both the curved and flat-space equations). Similarly, the obstruction we found to a higher-dimensional application of the Weyl doubling formula to our multi-Kerr-Schild single copy was in the inability to find the necessary scalar.

We also want to highlight the importance of the principal tensor as reviewed in \cite{Frolov:2017kze}.  Spacetimes
possessing a principal tensor are always algebraically special (type D).  If they are vacuum solutions, then they
also have aligned fields: electromagnetic fields with the same components as the principal tensor. These fields have $(D-\epsilon)/2$ free parameters; setting these charges proportional to the mass and $(D-2-\epsilon)/2$ NUT charges results in our single copy field. In some sense the principal tensor controls the form of the metric, so it is not too surprising that it is related so closely to the metric's single copy.  A natural extension of our work would be to include non-vacuum metrics which still possess a principal tensor.

Since the class of vacuum spacetimes which possess a principal tensor is the same Kerr-NUT-(A)dS class we consider here, any attempt to consider a broader set of vacuum solutions (e.g. bumpy black holes, black rings, etc) may require significant extension. 
Indeed, the argument for Weyl locality in position space given in \cite{Luna:2022dxo} indicates that
non-algebraically special spacetimes cannot have a local Weyl single copy, and spacetimes possessing a principal
tensor are always algebraically special.  Nevertheless, we hope that our work here may provide an approach to make further spacetimes at least perturbatively accessible.

We additionally hope our work may provide a framework to further study gravitational transformations and dualities in the double copy \cite{Bahjat-Abbas:2020cyb,Arkani-Hamed:2019ymq,Alawadhi:2019urr,Banerjee:2019saj,Huang:2019cja,Emond:2020lwi}, since it includes the full class of Kerr-NUT-(A)dS spacetimes.  Similarly, perhaps the principal tensor can be studied in the single-copy picture, just as NUT charges, geodesics, and Ricci flow have been \cite{Alfonsi:2020lub,Alawadhi:2022gwy,Gonzo:2021drq,Alawadhi:2021uie}.  Finally, for, the Kerr-NUT solutions in our family, we would be interested to see if the Weyl double copy in general dimensions can be understood asymptotically, as at null infinity in four dimensions \cite{Godazgar:2021iae,Luna:2016due,Gonzo:2022tjm,Chacon:2021hfe,Adamo:2021dfg,Campiglia:2021srh}.

\acknowledgments

The authors thank Damien Easson, Andres Luna, Tucker Manton, and Silvia Nagy for helpful discussions. The authors are supported by the U.S. Department of Energy under grant number DE-SC0019470 and by the Heising-Simons Foundation “Observational Signatures of Quantum Gravity” collaboration grant 2021-2818.

\begin{appendices}
\addtocontents{toc}{\protect\setcounter{tocdepth}{1}}

\section{Little group structure of type D Weyl spinor} \label{SpinorConventions} 
This brief appendix is a companion to section \ref{sec:ReviewSpinors5D} and gives details of calculations that
use the spinor set-up used in this paper. As an illustrative exercise in the spinor conventions used, we will lay out an argument for why
only $\Psi^{(2)} \neq 0$ for a type D spacetime.

First, we describe how the spinor basis $(k^{A}_{a}, n^{A}_{a})$ can be used to give back members of the null
frame \eqref{NullFrameMetric}. Effectively, this inverts the definition of the basis \eqref{SpinorBasisDefn} in
terms of the null frame. The vectors $k^{\mu}$ and $n^{\mu}$ can be obtained by contracting two copies of
$k^{A}_{a}$ and $n^{A}_{a}$ respectively with gamma matrices
\begin{equation} \label{GammaContractionsPartOne} 
    \begin{split}
        k_{a}^{A} \gamma^{\mu}_{AB} k_{b}^{B} &= - \sqrt{2} k^{\mu} \epsilon_{ab}, \\
        n_{a}^{A} \gamma^{\mu}_{AB} n_{b}^{B} &= - \sqrt{2} n^{\mu} \epsilon_{ab}.
    \end{split}
\end{equation}
The remaining contractions of $\gamma^{\mu}$ with the spinor basis produces the remaining members of the null frame
\begin{equation} \label{GammaContractionsPartTwo} 
    \begin{split}
        k_{a}^{A} \gamma^{\mu}_{AB} n^{B}_{b} &= - n_{a}^{A} \gamma^{\mu}_{AB} k^{B}_{b} \\
                                              &= 
                                              \begin{pmatrix}
                                                  \sqrt{2} m^{\mu} & i e_{0}^{\mu} \\
                                                  i e_{0}^{\mu} & \sqrt{2}  \overline{m}^{\mu}
                                              \end{pmatrix},
    \end{split}
\end{equation}
which, following \cite{Monteiro:2018xev}, we dub the polarization tensor $\varepsilon^{\mu}_{ab}$.

Next, we explain how to compute the components of the spinor Lorentz generators
in the $(k^{A}_{a}, n^{A}_{a})$ basis. Take, as an example, the contraction
$\sigma^{\mu \nu}_{AB} k^{A}_{a} k^{B}_{b}$. Noting the definition of the Lorentz generators in terms of gamma
matrices \eqref{SpinorLorentzGeneratorsDefn}, we have
\begin{equation} \label{SigmaComponentsStepOne} 
        \sigma_{AB}^{\mu \nu} k^{A}_{a} k^{a}_{b} = \gamma^{[\mu}_{AE} \gamma^{\nu]}_{FB} \Omega^{EF}
        k^{A}_{a} k^{B}_{b}.
\end{equation}
The normalization of the spinor basis \eqref{SpinorBasisNormalization} is equivalent to the following
expansion of $\Omega^{AB}$ 
\begin{equation} \label{OmegaExpansion} 
    \Omega^{AB} = \epsilon^{ab}\left(k^{A}_{a} n^{B}_{b} + n^{A}_{a} k^{B}_{b}\right).
\end{equation}
Substituting this back in \eqref{SigmaComponentsStepOne}, and keeping in mind \eqref{GammaContractionsPartOne}
and \eqref{GammaContractionsPartTwo} we obtain
\begin{equation} \label{SigmaComponentsPartOne} 
    \begin{split}
        \sigma_{AB}^{\mu \nu} k^{A}_{a} k^{B}_{b} &= \epsilon^{cd} \gamma^{[\mu}_{AE} \gamma^{\nu]}_{FB}
        \left(k^{E}_{c} n^{F}_{d} + n^{E}_{c} k^{F}_{d}\right) k^{A}_{a} k^{B}_{b} \\
                                                  &= \epsilon^{cd} \left(k^{A}_{a} \gamma^{[\mu|}_{AE}
                                                  k^{E}_{c} n^{F}_{d} \gamma^{|\nu]}_{FB} k^{B}_{b} +
                                              k^{A}_{a} \gamma^{[\mu |}_{AE} n^{E}_{c} k^{F}_{d} \gamma^{|
                                          \nu]}_{FB} k^{B}_{b} \right) \\
                                                  &= \epsilon^{cd} \left( \epsilon_{ac} (- \sqrt{2}
                                                  k^{[\mu})(-\varepsilon^{\nu]}_{bd}) +
                                          (\varepsilon^{[\mu}_{ac}) ( - \sqrt{2}
                                  k^{\nu]}\epsilon_{db})\right) \\
                                                  &= 2\sqrt{2} k^{[\mu} \varepsilon^{\nu]}_{ab}.
   \end{split}
\end{equation}
Similarly the other components are
\begin{equation} \label{SigmaComponentsPartTwo} 
    \begin{split}
        \sigma_{AB}^{\mu \nu} k^{A}_{a} n^{B}_{b} &= 2 k^{[\mu}n^{\nu]} \epsilon_{ab} +
        \varepsilon^{[\mu}_{ac} \varepsilon^{\nu]}_{bd} \epsilon^{cd}, \\
        \sigma_{AB}^{\mu \nu} n^{A}_{a} k^{A}_{a} &= 2 n^{[\mu} k^{\nu]} \epsilon_{ab} +
        \varepsilon^{[\mu}_{ac} \varepsilon^{\nu]}_{bd} \epsilon^{cd}, \\
        \sigma_{AB}^{\mu \nu} n^{A}_{a} n^{B}_{b} &= - 2\sqrt{2} n^{[\mu} \varepsilon^{\nu]}_{ab}.
    \end{split}
\end{equation}

We are now prepared to tackle the little group structure of a type D Weyl spinor. Inverting the expansion of
$\Psi_{ABCD}$ \eqref{WeylLittleGroupObjects}, we obtain
\begin{equation} \label{} 
    \begin{split}
        \Psi^{(0)}_{abcd} &= \Psi_{ABCD} n^{A}_{a} n^{B}_{b} n^{C}_{c} n^{D}_{d}, \\
        \Psi^{(1)}_{abcd} &= \Psi_{ABCD} n^{a}_{a} n^{b}_{b} n^{c}_{c} k^{d}_{d}, \\
        \Psi^{(2)}_{abcd} &= \Psi_{ABCD} n^{a}_{a} n^{b}_{b} k^{c}_{c} k^{d}_{d}, \\
        \Psi^{(3)}_{abcd} &= \Psi_{ABCD} n^{a}_{a} k^{b}_{b} k^{c}_{c} k^{d}_{d}, \\
        \Psi^{(4)}_{abcd} &= \Psi_{ABCD} k^{a}_{a} k^{b}_{b} k^{c}_{c} k^{d}_{d}.
    \end{split}
\end{equation}
Next, we rewrite $\Psi_{ABCD}$ in terms of the Weyl tensor $W_{\mu \nu \rho \lambda}$ using \eqref{WeylSpinorDefn} 
\begin{equation} \label{} 
    \begin{split}
        \Psi^{(0)}_{abcd} &= W_{\mu \nu \rho \lambda} \sigma^{\mu \nu}_{AB} \sigma^{\rho \lambda}_{CD} n^{A}_{a} n^{B}_{b} n^{C}_{c} n^{D}_{d}, \\
        \Psi^{(1)}_{abcd} &= W_{\mu \nu \rho \lambda} \sigma^{\mu \nu}_{AB} \sigma^{\rho \lambda}_{CD} n^{a}_{a} n^{b}_{b} n^{c}_{c} k^{d}_{d}, \\
        \Psi^{(2)}_{abcd} &= W_{\mu \nu \rho \lambda} \sigma^{\mu \nu}_{AB} \sigma^{\rho \lambda}_{CD} n^{a}_{a} n^{b}_{b} k^{c}_{c} k^{d}_{d}, \\
        \Psi^{(3)}_{abcd} &= W_{\mu \nu \rho \lambda} \sigma^{\mu \nu}_{AB} \sigma^{\rho \lambda}_{CD} n^{a}_{a} k^{b}_{b} k^{c}_{c} k^{d}_{d}, \\
        \Psi^{(4)}_{abcd} &= W_{\mu \nu \rho \lambda} \sigma^{\mu \nu}_{AB} \sigma^{\rho \lambda}_{CD} k^{a}_{a} k^{b}_{b} k^{c}_{c} k^{d}_{d}.
    \end{split}
\end{equation}
Finally, substituting the components of $\sigma^{\mu \nu}_{AB}$ we just found in
\eqref{SigmaComponentsStepOne}, \eqref{SigmaComponentsPartTwo},
\begin{equation} \label{} 
    \begin{split}
        \Psi^{(0)}_{abcd} &= 8 W_{\mu \nu \rho \lambda} k^{\mu} \varepsilon^{\nu}_{ab} k^{\rho}
        \varepsilon^{\lambda}_{cd}, \\
        \Psi^{(1)}_{abcd} &= 2\sqrt{2} W_{\mu \nu \rho \lambda} k^{[\mu} \varepsilon^{\nu]}_{ab} \left( 2k^{[\rho}
        n^{\lambda]} \epsilon_{cd} + \epsilon^{gh} \varepsilon^{[\rho}_{cg}
    \varepsilon^{\lambda]}_{hd}\right), \\
            \Psi^{(2)}_{abcd} &= -8 W_{\mu \nu \rho \lambda} k^{[\mu} \varepsilon^{\nu]}_{ab} n^{[\rho}
            \varepsilon^{\lambda]}_{cd}, \\
            \Psi^{(3)}_{abcd} &= -2 \sqrt{2}  W_{\mu \nu \rho \lambda} \left( 2k^{[\mu} n^{\nu]} \epsilon_{ab} +
            \epsilon^{gh} \varepsilon^{[\mu}_{ag} \varepsilon^{\nu]}_{hb}\right) n^{[\rho}
            \varepsilon^{\lambda]}_{cd}, \\
                \Psi^{(4)}_{abcd} &= 8 W_{\mu \nu \rho \lambda} n^{\mu} \varepsilon^{\nu}_{ab} n^{\rho}
                \varepsilon^{\lambda}_{cd}.
    \end{split}
\end{equation}
Now, remembering that $\varepsilon^{\mu}_{ab}$ is orthogonal to both $k^{\mu}$ and $n^{\mu}$, a comparison with
\eqref{TypeDConditions} shows that with a properly chosen null frame $\Psi^{(0)}, \Psi^{(1)}, \Psi^{(3)}$ and
$\Psi^{(4)}$ all vanish for a type D spacetime.

\end{appendices}

\bibliographystyle{jhep.bst}
\bibliography{5DDoubleCopy} 

\end{document}